%% file: main.tex
\documentclass[a4paper,conference]{IEEEtran}
\IEEEoverridecommandlockouts
\usepackage{fancyhdr}
\usepackage{float}
\usepackage{graphicx} 
\usepackage[utf8]{inputenc}
\usepackage{todonotes}
\usepackage{comment}
\usepackage{hyperref}
\usepackage{amsmath}  
\usepackage{amssymb}  
\usepackage{colortbl} 
\input{symbols_def}

\begin{document}

\title{A Single Subject Machine Learning Based Classification of Motor Imagery EEGs

\thanks{$^{1}$ Department of Electrical Electronic and Computer Science Engineering \\ $^{2}$ Istituto di Scienze Applicate e Sistemi Intelligenti "E. Caianiello", $^{3}$Consiglio Nazionale delle Ricerche.\\
This work has been funded by the MUR Program  PRIN 2022, project HOME4.0 (E53D23000510006).
}}

\author{\IEEEauthorblockN{1\textsuperscript{st} Dario Sanalitro}
\IEEEauthorblockA{\textit{DIEEI$^{1}$} \\
\textit{University of Catania}\\
Catania, Italy \\
dario.sanalitro@unict.it}\\
\IEEEauthorblockN{4\textsuperscript{th} Pasquale Memmolo}
\IEEEauthorblockA{\textit{ISASI$^{2}$} \\
\textit{CNR$^{3}$}\\
Pozzuoli, Italy \\
pasquale.memmolo@isasi.cnr.it}\\
\and
\IEEEauthorblockN{2\textsuperscript{nd} Marco Finocchiaro}
\IEEEauthorblockA{\textit{DIEEI$^{1}$} \\
\textit{University of Catania}\\
Catania, Italy \\
marco.finocchiaro@phd.unict.it}
\and
\IEEEauthorblockN{3\textsuperscript{rd} Emanuela Cutuli}
\IEEEauthorblockA{\textit{DIEEI$^{1}$} \\
\textit{University of Catania}\\
Catania, Italy \\
emanuela.cutuli@phd.unict.it}\\
\IEEEauthorblockN{5\textsuperscript{th} Maide Bucolo}
\IEEEauthorblockA{\textit{DIEEI$^{1}$} \\
\textit{University of Catania}\\
Catania, Italy \\
maide.bucolo@unict.it}
}

\maketitle
\thispagestyle{fancy}
\begin{abstract}
Motor Imagery-Based Brain-Computer Interfaces (MI-BCIs) are systems that detect and interpret brain activity patterns linked to the mental visualization of movement, and then translate these into instructions for controlling external robotic or domotic devices. 
Such devices have the potential to be useful in a broad variety of applications.
While implementing a system that would help individuals restore some freedom levels, the interpretation of (Electroencephalography) EEG data remains a complex and unsolved problem.
In the literature, the classification of left and right imagined movements has been extensively studied. This study introduces a novel pipeline that makes use of machine learning techniques for classifying MI EEG data. The entire framework is capable of accurately categorizing left and imagined motions, as well as rest phases, for a set of 52 subjects who performed a MI task. We trained a \textit{within subject model} on
each individual subject. The methodology has been offline evaluated and compared to four studies that are currently the state-of-the-art regarding the specified dataset.  The results show that our proposed framework could be used with MI-BCI systems in light of its failsafe classification performances, i.e. 99.5\% in accuracy.
	
\end{abstract}


%
\IEEEpeerreviewmaketitle

\section{Introduction}
The human brain has been the subject of extensive study for many years due to its intriguing nature as a dynamic and sophisticated organ. One interesting area that has gained significant attention over the years is the possibility of creating a direct connection between the brain and an external device bypassing the body’s more typical pathways of nerves
and muscles~\cite{2012-Shi}. These systems are well-known in the literature as Brain Computer Interfaces (BCIs).
A huge number of studies have explored the potential of BCI systems in various applications, including rehabilitation~\cite{2021-Rob}, navigation~\cite{2021-Zha}, environmental control~\cite{2023-CarSanMicBusBuc}, gaming and entertainment~\cite{2019-Aro}. 
However, there are still many areas that have not been fully explored where these systems might possibly be used. Among them, manipulating single and multi-robot systems or performing tasks in interaction with the external environments due to their capabilities of manipulating objects~\cite{2019-RylMusPieCatAntCacBicFra, 2020-UmiSanOriFra, 2022-SanTogJimCorFra}.

Various types of task-related information can be extracted from brain waves, depending on the chosen experimental method and the neurophysiological activation pattern. Examples include evoked potentials (EPs), event-related potentials (ERP)~\cite{1999-Pfu}, sensorimotor rhythms as Motor Imagery (MI)~\cite{2014-Yua}.
The latter is a cognitive process in which the user mentally visualizes motor movements without physically activating any muscle or peripheral nerves.

\begin{figure}[t]
	\includegraphics[width=\columnwidth]{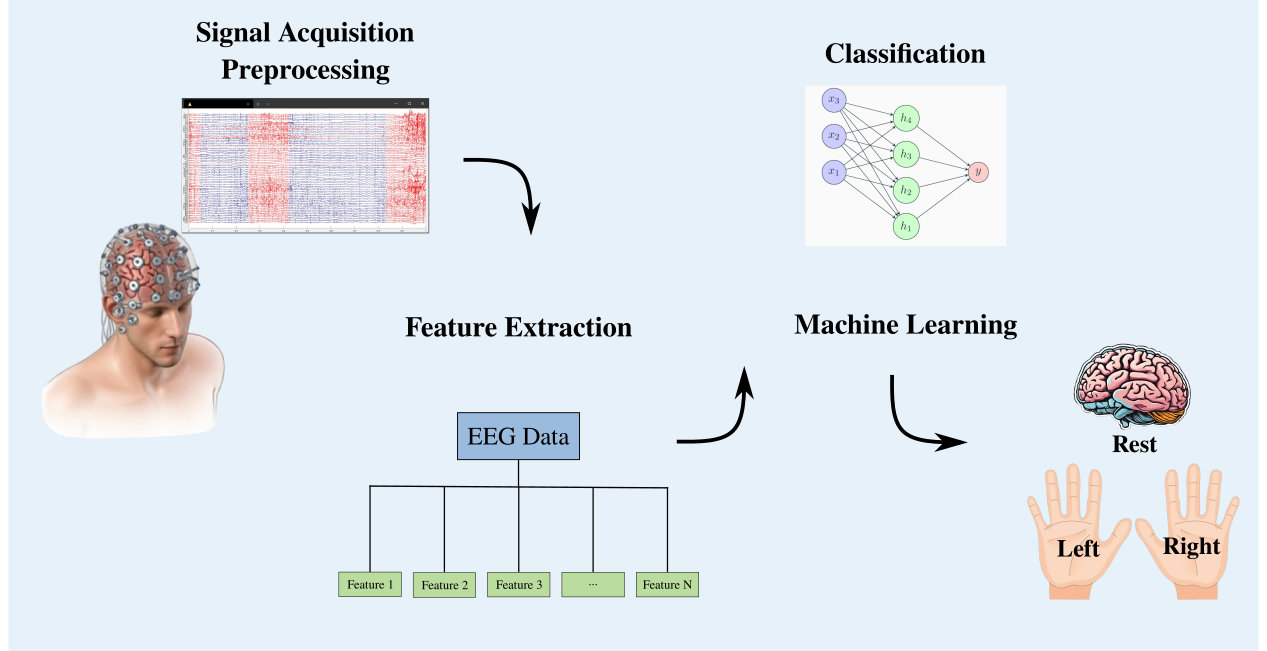}
	\caption{Overview of the EEG data processing for MI classification of left, right hand movements and rest phase.}
\end{figure} 	

When assessing brain activity, most of the cutting-edge methods for MI decoding take EEGs into account.
Despite being a reasonable compromise in terms of cost, portability, and temporal precision, EEGs have a limited signal-to-noise ratio and limited spatial resolution.
Therefore, advanced signal processing techniques are necessary to extract task-specific brain-related characteristics.
The main actions that human beings may conceive, that have been reported in the literature and classified through EEG data are hand motions, arm movements, leg movements, and tongue movements~\cite{2008-MorBaiFurLinHal}. 

Extensive literature research has been conducted on the classification of imagined movements. Classical approaches aim to highlight spatial features linked to movement imagination. 
As a matter of fact, when subjects imagine left or right hand, arm, or foot movements, the data show a lateralization, particularly when the frequency EEG band [7-30 Hz] is considered.
Examples of spatial filtering algorithms, such as Common Spatial Patterns (CSP)~\cite{2007-Dor} are used to detect the lateralization on subjects who are able to modulate their brain activity in a way suitable for the lateralization detection. 
Such a result is achieved by maximizing the variance for one class while minimizing the variance for the other classes~\cite{2018-NamNijLot}.
Regularized Common Spatial Patterns (RCSP)~\cite{2010-Lot}, Wavelet Common Spatial Patterns (WCSP), Common Spatio-Spectral Patterns (CSSP) or Common
 sparse spectral spatial pattern (CSSSP) are all methodologies that are effective variants of the original algorithm~\cite{2011-MouMalFitLit}.

Other approaches perform the classification by employing linear classifiers such as linear discriminant analysis (LDA) or support vector machine (SVM), respectively used for controlling a domotic system in~\cite{2023-CarSanMicBusBuc} and for a comparison of subject-independent and subject-specific EEG-based BCI in~\cite{2023-DosSanFra}. Artificial Neural Networks, such as the Multi-Layer Perceptron (MLP), represent another class of classifiers which have been largely used for MI classification. Particularly, in~\cite{2022-ShaKimGup} an MLP has been tested over 4 subjects achieving high accuracy while classifying left and right hand imagined movements.

In the recent years, there has been a rise in the use of Deep Learning techniques as  Convolutional Neural Networks (CNN) thanks to their ability of learning features directly from large datasets showing promising results in EEG MI classification.  Among them,~\cite{2017-TanLiSun} uses CNN for single trials MI-EEG classification,~\cite{2017-SchSprFieGlaEggTanHutBurBall} describes deep Convolutional Networks for decoding imagined or executed tasks from raw EEG, ~\cite{2022-TibLeeAli} exploit CNNs for improving BCI performances. 
One of the most used network~\cite{2018-LawSolWayGorSteHunLan}, EEGNet, a compact convolutional neural network for EEG-based BCIs, has shown its optimal performances, both for within-subject across various BCI paradigms. 
In this work, we propose a pipeline and a feature extraction that shows capabilities in competing with the best current state-of-the-art approaches in classifying motor imagery tasks associated with imagined motions of right and left hand. We performed a comparison of our results with those reported in four recent studies \cite{2017-HohAhnAhnKwoMooJun}, \cite{2019-Kum}, \cite{2021-Sad}, \cite{2022-Sad}, which analyzed the same dataset. They served as our point of comparison, considering that they represent the current state-of-the-art in general and specifically for this particular dataset.
Hence, the main contributions of this work can be summarized in \textit{i)} the introduction of a significantly simpler processing pipeline in comparison with other studies, \textit{ii)} the improved classification performance in terms of accuracy, outperforming the existing approaches studying this dataset \textit{iii})  effective classification of person-dependent Sensory Motor Rhythms (SMRs) on large-scale datasets (52 subjects) in contrast with state-of-the-art approaches that analyze datasets considering a limited number of subjects. The results of the classification tests of a motor imagery task show that the acquired accuracy rates never go beyond 99.5\%. 
The paper is organized as follows. In \sect\ref{sec:MM}, the dataset and the experimental protocol are described. In \sect\ref{sec:Method}, we present the proposed pipeline while in \sect\ref{sec:Results}, we show the obtained results.



\section{Materials}\label{sec:MM}

\subsubsection*{Subjects}
The study was conducted on 52 subjects and the data was taken from~\cite{2017-HohAhnAhnKwoMooJun}.
The authors used 52 volunteers, 19 of whom were female, with an average age of 24.8 ± 3.86 years. Out of fifty participants, two were both-handed and the others were right-handed.
\subsubsection*{Data Recording}
\begin{figure}[t]
	\centering
	\includegraphics[width=0.6\columnwidth]{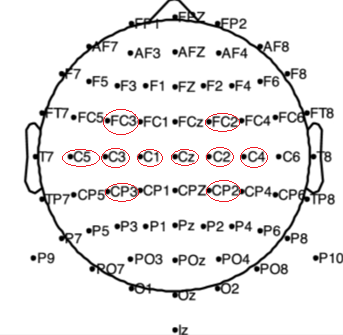}
	\caption{Number of electrodes in the dataset}
	\label{fig:selection}
\end{figure}

The EEG signals were recorded using a Biosemi ActiveTwo 64-channel montage based on the worldwide 10-10 system at 512 Hz sampling rates, as illustrated in \fig\ref{fig:selection}. In addition, the authors used BCI2000 for EEG data collection and instructions presentation.
\subsubsection*{Experiment}
\begin{figure}[t]
	\centering
	\includegraphics[width=1\columnwidth]{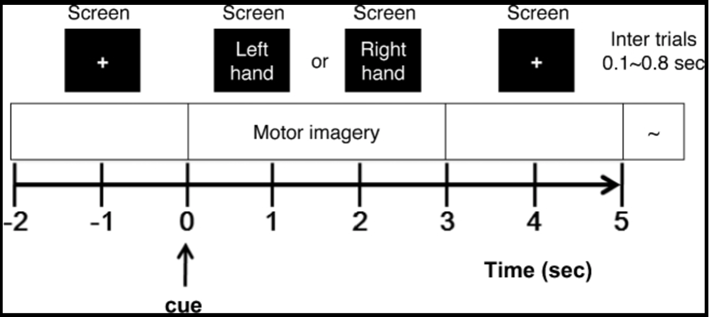}
	\caption{registration steps epoch}
	\label{fig:Epoch}
\end{figure}

The protocol is shown in \fig\ref{fig:Epoch}.
Before each trial began, the subject was instructed to wait two seconds for the monitor to display a blank screen with a fixation cross. 
Then, for three seconds, one of the two instructions, either left or right, was displayed. Subjects were asked to visualize moving their hand in response to the displayed instruction within this time interval. Upon the return of the blank screen, a random interval of 0.1 to 0.8 seconds was provided to relax. A total of 40 iterations of these steps were carried out for each run.
For our purposes, the three seconds of instructions and the resting state were considered. 

\section{Methods}\label{sec:Method}

To develop a complete framework that could be used in the future for real-time purposes, we experimented with several processing pipelines and ultimately adopted a structure inspired by the approach proposed in~\cite{2018-MohYus}.  A simplified pipeline obtained by removing one of their preprocessing step, i.e. the Empirical Mode Decomposition, results in a more efficient framework in terms of computational time. Thus, our model was trained by first pre-processing the data, extracting two main features, i.e. Energy and Instantaneous Entropy. Then, we used state-of-the-art algorithms to classify the brain lateralization. We first selected 10 channels out of the 64 available, to better examine the motor cortex phasic activations with a focus on identifying SMRs that are predominantly found over the central scalp areas. Red circles in \fig\ref{fig:selection} shows the selected channels for our analysis.
A block diagram representing all these stages is instead shown in Fig. ~\ref{fig:framework}.

\input{./img/flow-chart.tex}

\subsection{Preprocessing}
Pre-processing steps included channels reduction, artifacts removal, frequency and spatial filtering. The channels reduction involved considering 10 channels out of the 64 channels available to specifically target the primary motor cortex involved in motor imagination. Afterwards, a statistical outlier detection has been implemented. The main objective was the discrimination of the outlier signal and its subsequent removal from the data. To perform such a removal, we computed the mean $\mu$ and standard deviation $\sigma$ across all epochs and for each subject. If the mean ($\mu_i$) in each epoch was beyond the range $\mu_i \notin [\mu-3\sigma;\mu+3\sigma]$, then it was considered as an outlier and removed from the data.

With the aim of studying the phasic activations in the motor cortex, we narrowed the data frequency range using a bandpass filter.
In particular, the data was filtered in the 8–30 Hz frequency range by employing a Butterworth bandpass filter of order 30, sampling at 512 Hz.
As last stage of our pre-pocessing pipeline, we applied a well-known Common Average Reference (CAR) spatial filter to average the voltage across all the selected electrodes. The methodology computes the average voltage $V(t)$ that is subtracted from the voltage measured at each individual electrode as follows
\begin{equation}
V_{CAR}(t)=V(t) - \frac{1}{N} \sum_{k=1}^{N}V_k(t)
\end{equation}
where $V_k(t)$ is the raw voltage at the electrode $k$ and $N$ is the considered electrodes number.
Through the process of subtraction, CAR effectively mitigates the impact of shared noise on all electrodes. Thus, a more distinct brain activity patterns can be analyzed.

\subsection{Feature Extraction}
Data transformation is the procedure of converting unprocessed data into numerical characteristics that can be analyzed while maintaining the integrity of the original dataset. According to~\cite{2018-MohYus}, two metrics have been selected, i.e. Energy and Instantaneous Spectral Entropy (ISE).

\subsubsection{Energy}
The Energy has been used to quantify the signal strength over the specific time frame of the MI task, for each epoch, we computed
\begin{equation}
	E_s = \sum_{n=-N}^{N} |x(n)|^2 
\end{equation}
where $x(n)$ are the signal samples and $N$ is the number of samples.
\subsubsection{Istantaneous Spectral Entropy (ISE)}
The ISE is proposed as a metric for capturing the complexity and unpredictability of the EEG signals. The concept is based on the notion that the higher the level of activity in a particular brain region or frequency band, the more complex the processed signal becomes, leading to higher entropy levels.
Specifically, it is derived by the power spectrum. Thus, given a signal $x(n)$, the power spectrum is
\begin{equation}
	S(f)=|X(f)|^2
\end{equation}
where $X(f)$ is the Discrete Fourier Transform of $x(n)$.
The probability distribution is then
\begin{equation}
	P(f)=\frac{S(f)}{\sum_{k}^{M}S(k)}
\end{equation}
where $M$ represents the total number of discrete frequencies.
Therefore, for a given probability distribution $P(f)$, the Shannon Spectral Entropy (SE) can be defined as
\begin{equation}
	H = -\sum_{k=1}^{M}P(k)\log_{2}P(k)
	\label{eqn:SE}
\end{equation}
Similarly, the probability distribution at time $t$ can be defined as
\begin{equation}
P(t,f)=\frac{\sum_{t}^{}S(t,f)}{\sum_{k}^{M}\sum_{t}^{}S(t,k)}
\end{equation}
and therefore the ISE becomes
\begin{equation}
H(t) = -\sum_{k=1}^{M}P(t,k)\log_{2}P(t,k)
\end{equation}
similarly to \eqn\ref{eqn:SE}.

\subsection{Classification}
In this section, we show the trained classifiers that were used for our analyses. The top candidates in terms of accuracy and training computational performance have been selected. These include Quadratic Discriminant Analysis (QDA), Fine K-Nearest Neighbor (KNN), Cosine KNN and Wide Neural Network (NN). Some details about them will be provided here. The interested reader is referred to more advanced textbooks for a better understanding.

\subsubsection{Quadratic Discriminant Analysis (QDA)}
Quadratic Discriminant Analysis (QDA) is a probabilistic classification method that given a feature vector $\vect{x}$, classifies it into class $C_k$ based on the posterior probability.
The posterior probability that a point $x$ belongs to class $C_k$ is the product of the prior probability and the multivariate normal density. The density function of the multivariate normal distribution with a mean vector $\vect{\mu}_k \in \nR{1\times d}$ and a covariance matrix $\vect{\Sigma}_k \in \nR{d \times d}$ at a $\vect{x} \in \nR{1 \times d}$ is given by:

\newcommand{\class}{C_k}
\begin{equation}
P(\vect{x} | \class) = \frac{1}{(2\pi)^{d/2} |\vect{\Sigma}_{\class}|^{1/2}} 
\exp \left( -\frac{1}{2} (\vect{x} - \vect{\mu}_{\class}) \vect{\Sigma}_{\class}^{-1} (\vect{x} - \vect{\mu}_{\class})^T \right)
\end{equation}
where $|\vect{\Sigma}_k|$ is the determinant of $\vect{\Sigma}_k$ and $\vect{\Sigma}_k^{-1}$ is its inverse matrix.
If $P(C_k)$ represents the prior probability of class $C_k$, the posterior probability that an observation $x$ belongs to class $C_k$ is

\begin{equation}
\hat{P}(C_k | \vect{x}) = \frac{P(\vect{x} | C_k) P(C_k)}{P(\vect{x})}
\end{equation}

where $P(\vect{x})$  is a normalization constant.

The prediction is finally done by minimizing the expected classification cost

\begin{equation}
\hat{y} = \arg\min_{C_k} \sum_{i}^{k} \hat{P}(C_k|\vect{x})J(y|C_k)
\end{equation}
where $\hat{y}$ is the predicted classification, $k$ is the number of classes, $J(y|C_k)$ is the cost of classifying an observation as y when its true class is $C_k$.

\subsubsection{Fine K-Nearest Neighbor (KNN)}
K-Nearest Neighbor (KNN) is a classification method that assigns a class label based on the majority vote of its $K$ nearest neighbors~\cite{2013-Kra}. In Fine KNN, a small value of $K$ (in our case 1 neighbor) has been used to capture a finely detailed distinction between classes. Given a distance metric $d(\vect{x}_i, \vect{x}_j)$, in our case the euclidean, the class prediction $\hat{y}$ for a new data point $\vect{x}$ is determined by:

\begin{equation}
\hat{y} = \arg\max_{y \in C_k} \sum_{i \in \mathcal{N}_k(\vect{x})} \mathcal{I}(y_i = y)
\end{equation}

where $\mathcal{N}_k(\vect{x})$ denotes the set of $k$ nearest neighbors of $\vect{x}$, and $\mathcal{I}(\cdot)$ is the indicator function, that returns \textit{one}, if its argument is true and \textit{zero} otherwise. 
\subsubsection{Wide Neural Network (NN)}
The Wide Neural Network (NN) is a neural network that consists of an input layer, a first fully connected layer with 10 outputs. Then, there is a ReLU activation function which performs a threshold operation on each element of the input. Then, a fully connected layer with $K$ layers corresponding to the classes number is the input of a softmax function which provides the predicted classification scores according to 

\begin{equation}
	f(x_i) = \frac{e^{x_i}}{\sum_{j=1}^{K}e^{x_j}}
\end{equation} 
where $x_i$ is the input and $K$ is the number of classes. Such layer is then connected to the output and final layer which corresponds to the predicted class labels.
\subsection{Evaluation}
We computed common evaluation metrics to gain a comprehensive understanding of the model's classification performance of the MI data. Particularly relevant for us are accuracy, recall or sensitivity, specificity and F1 Score which are defined by
\begin{equation}
Accuracy = \frac{TP + TN}{TP + TN + FP + FN}
\end{equation}
\begin{equation}
Recall = \frac{TP}{TP + FN}
\end{equation}
\begin{equation}
Specificity = \frac{TN}{TN + FP}
\end{equation}
\begin{equation}
F_{1} = \frac{TP}{TP + \frac{1}{2}(FP+FN)}
\end{equation}
where $TP$ and $TN$ represent true positive and true negative, while $FP$ and $FN$ represent false positive and false negative.

\begin{figure}[t]
\centering
	\includegraphics[width=0.8\columnwidth]{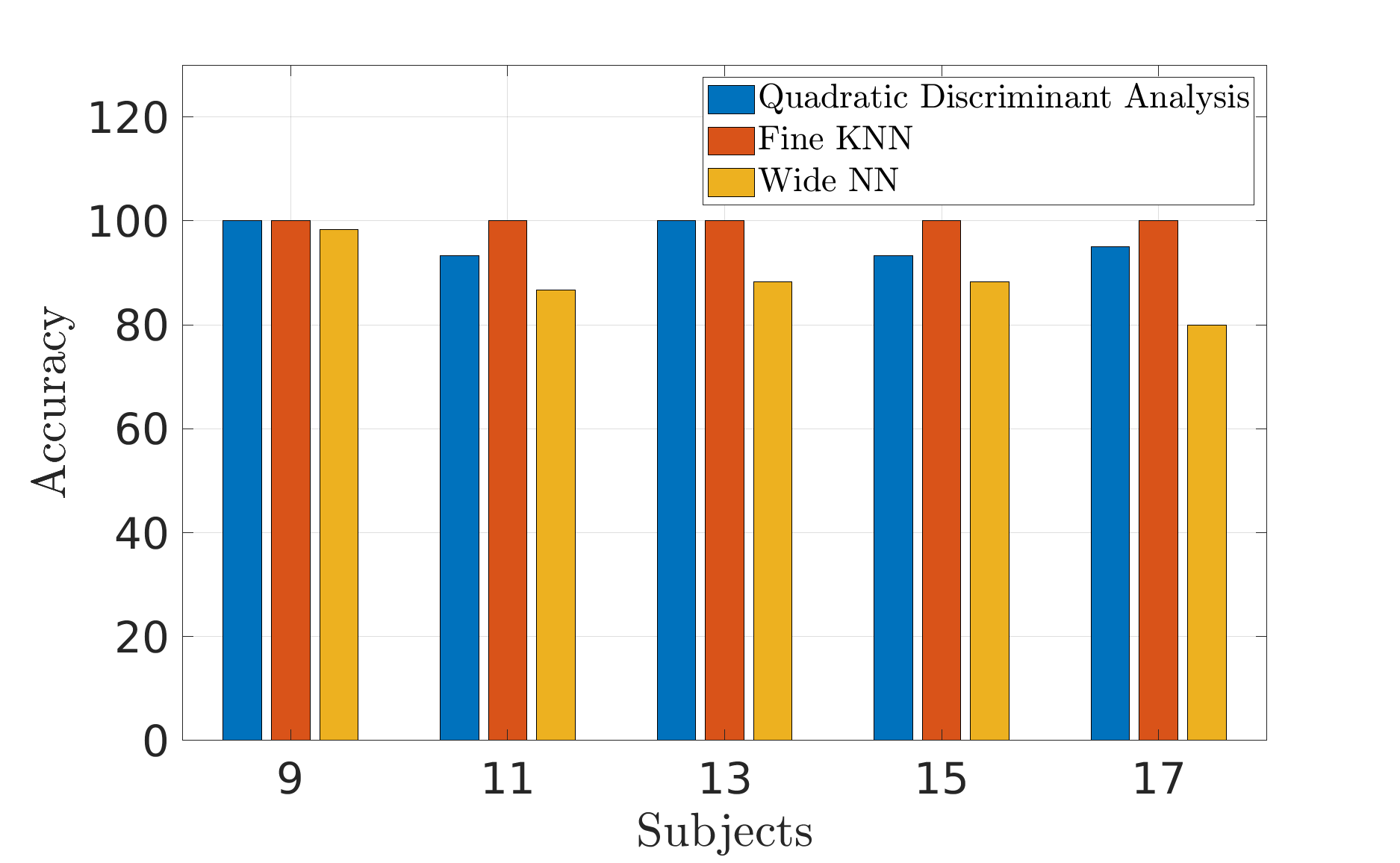}
	\caption{Accuracy comparison among the results of the four trained models: QDA, Cosine KNN, Fine KNN, Wide NN during the testing phase for the a subset of subjects participating in the study, i.e. S9, S11, S13, S15, S17. They were randomly selected.}
	\label{fig:accuracy}
\end{figure}

\begin{figure}[htbp]
	\centering
	\begin{subfigure}[b]{0.49\textwidth}
		\centering
		\includegraphics[width=\textwidth]{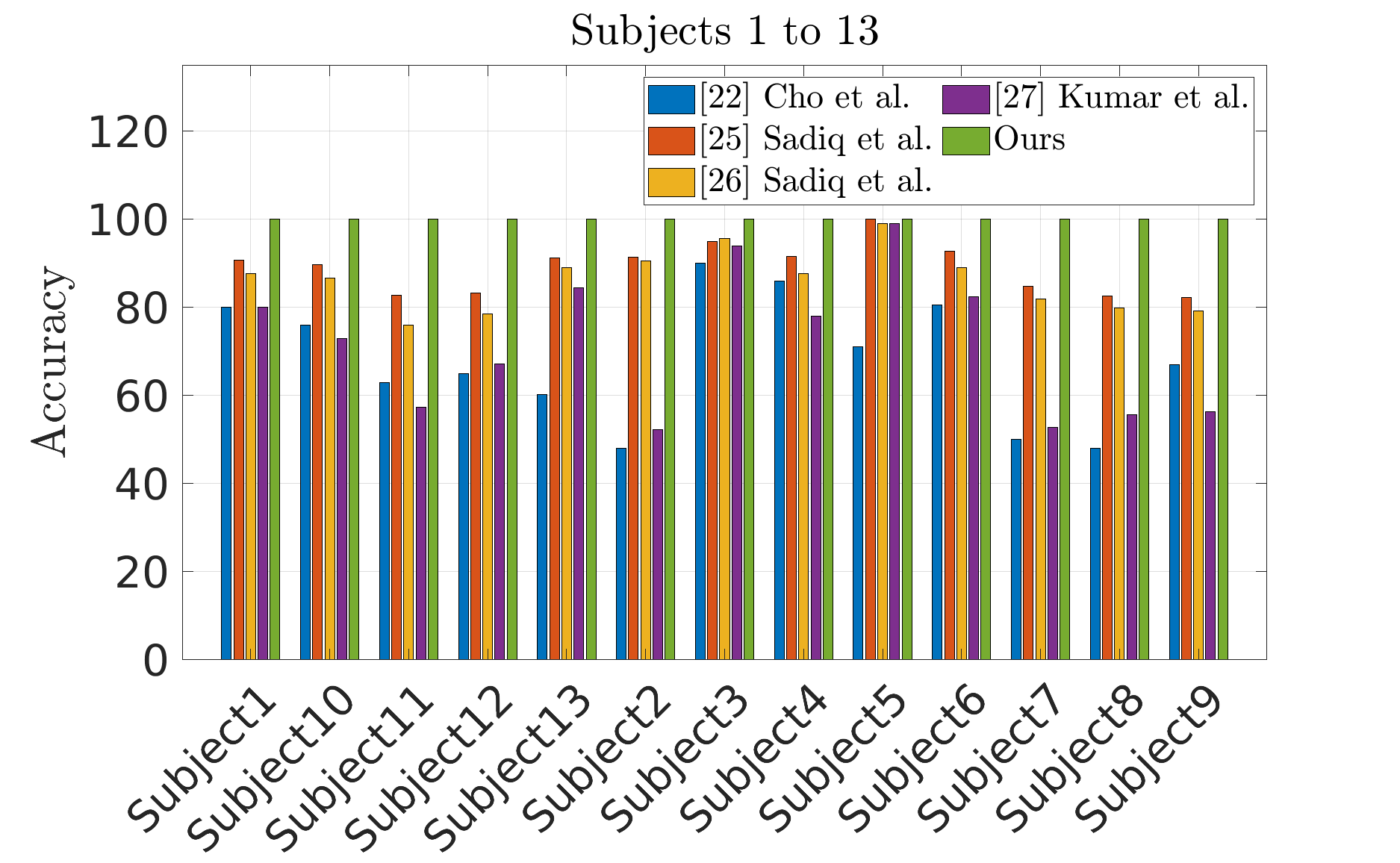}
	\end{subfigure}
	\begin{subfigure}[b]{0.49\textwidth}
		\centering
		\includegraphics[width=\textwidth]{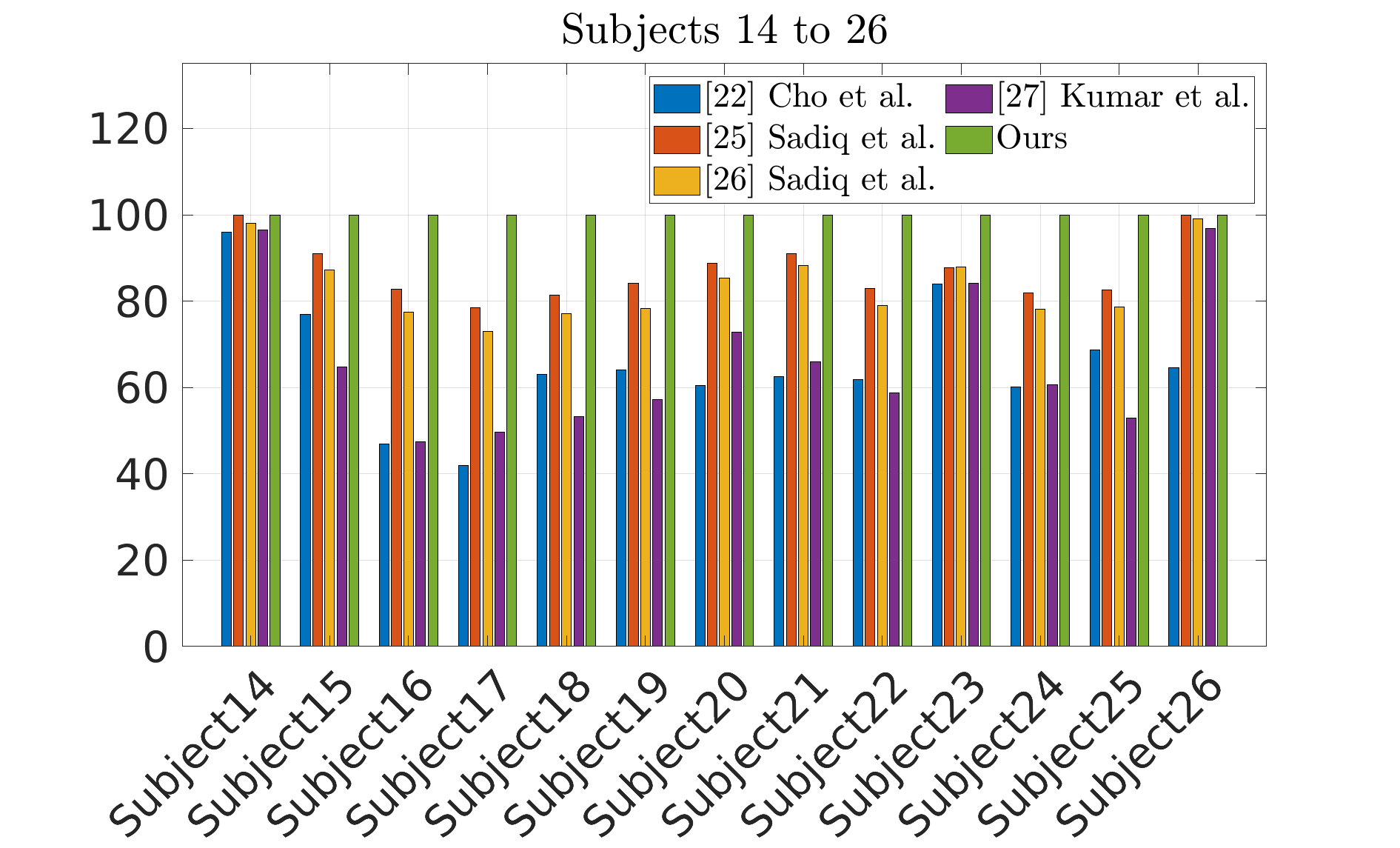}
	\end{subfigure}
	
	\begin{subfigure}[b]{0.49\textwidth}
		\centering
		\includegraphics[width=\textwidth]{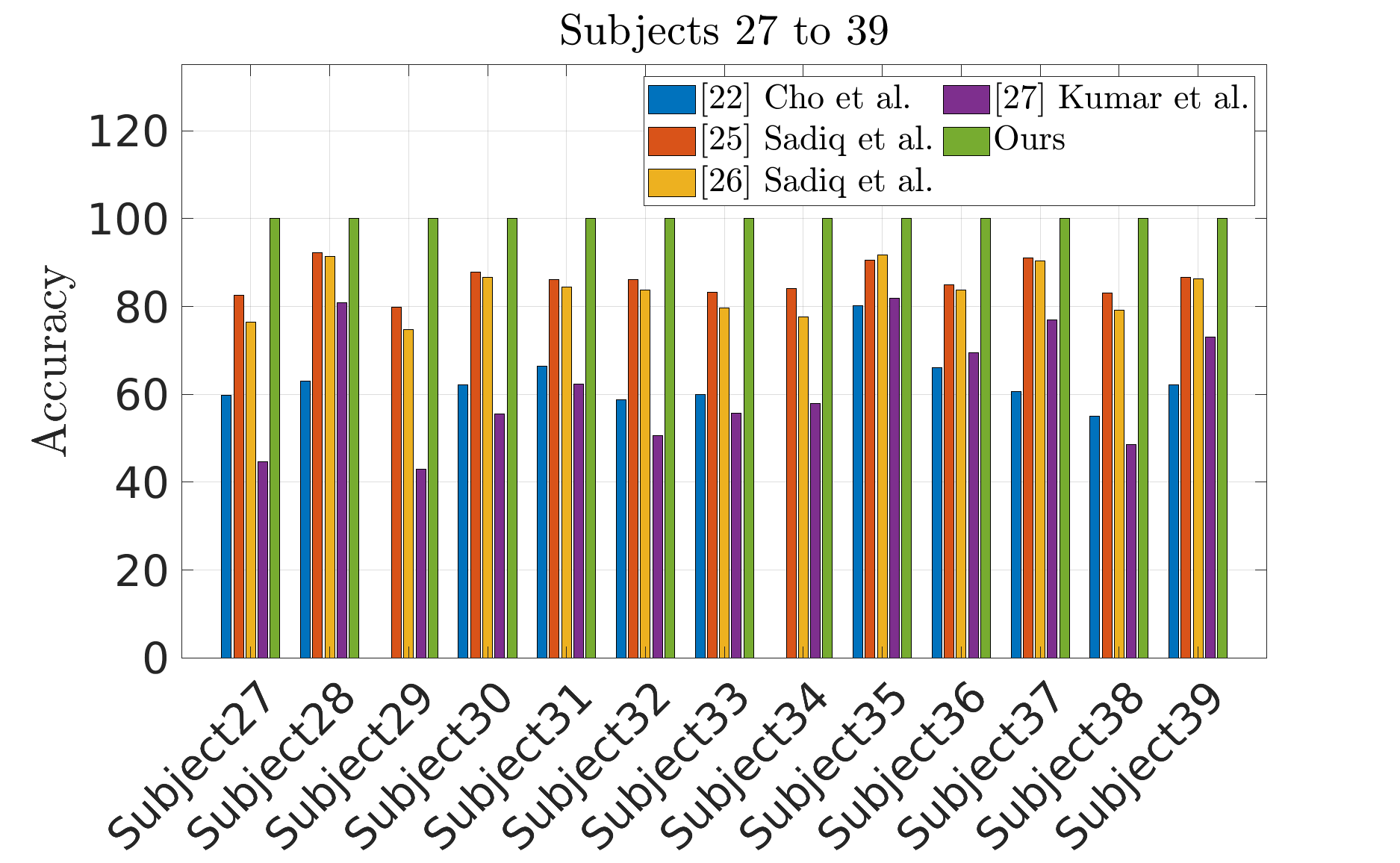}
	\end{subfigure}
	\begin{subfigure}[b]{0.49\textwidth}
		\centering
		\includegraphics[width=\textwidth]{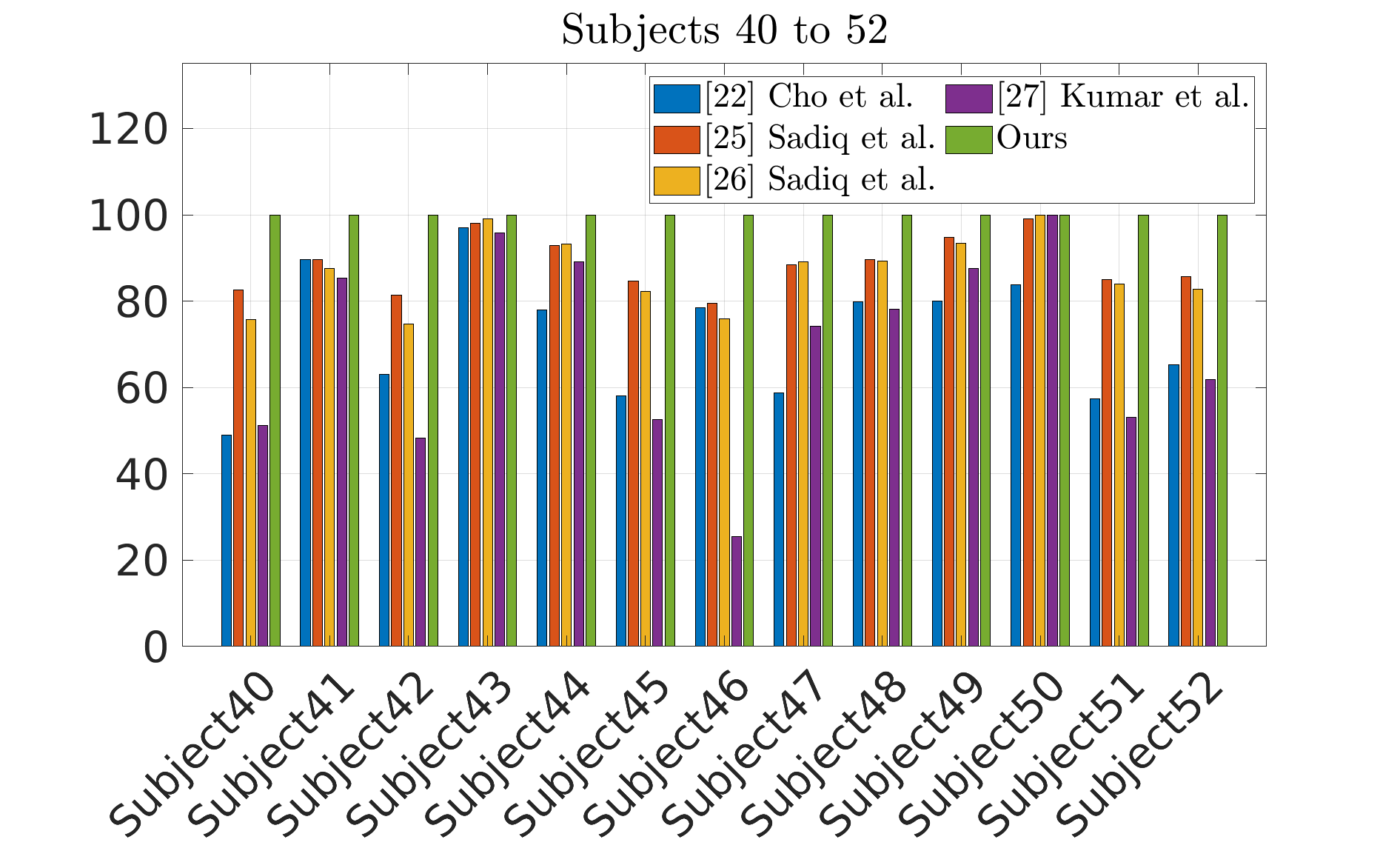}
	\end{subfigure}

	\caption{Comparison between the findings present in the state-of-the-art, specifically~\cite{2017-HohAhnAhnKwoMooJun, 2019-Kum, 2021-Sad, 2022-Sad} and ours. }
	\label{fig:accuracy_all}
\end{figure}
\section{Results}\label{sec:Results}
In this section, for the whole set of 52 subjects, we show the results coming from four well-known learned models in general classification tasks. 
In particular, we trained QDA, Fine-KNN, Cos-KNN, and Wide-NN since they outperformed the others in terms of validation and test accuracy, computation time and model size within this particular processing pipeline.
The classification learner from MATLAB has been used for our analysis, running on a portable PC equipped with Intel-Evo i7 processor and 16GB of RAM. 
In \tab\ref{tab:model_performance}, we summarize the training durations, validation and test accuracies, model sizes for the whole dataset. Even though QDA showed superior training computational efficiency (12.5 $\rm s$) and NN was better in terms of model size (256 $\rm kB$), the Fine KNN emerged as the most accurate model, achieving 99.6\% and 100\% accuracy in validation and testing, despite its larger model size which, in any case, remains constrained (30 $\rm MB$). Therefore, the whole analysis was conducted by using it as the employed model.
We trained it over the 100 trials available for each subject. 80\% of the data was given for the training and validation phases and 20\% was given for the test phase. The accuracy for each subject was then computed.  \fig\ref{fig:accuracy} shows the results for a subset of six subjects, i.e. S9, S11, S13, S15, S17, S19, obtained by employing the aforementioned models. Such a result shows the better performances of Fine KNN over the other models in a subject-specific, consistently exceeding 99.5\%. 
In addition, \fig\ref{fig:accuracy_all} shows the accuracies obtained for all the 52 subjects in comparison with the results presented in~\cite{2017-HohAhnAhnKwoMooJun, 2019-Kum, 2021-Sad, 2022-Sad} where the same dataset has been analyzed. The comparison shows our better performances in every single case with respect to the other works in all tested scenarios.
The average results expressed in terms of average accuracy, precision, recall and $F_1$, for each of work, are presented in \tab\ref{tab:avg-res} . In particular, the average accuracy, recall, specificity and F$_1$ score are 0.995,0.9986,0.9983,0.9984, demonstrating once again better performances.
All these evidences demonstrate that the pipeline described in this work can effectively classify person-dependent SMRs using a large and public dataset.

\begin{table}[t]
	\centering
	\begin{tabular}{|l|c|c|c|c|}
		\hline
		\centering\textbf{Model} & \multicolumn{1}{c|}{\textbf{Validation}} & \multicolumn{1}{c|}{\textbf{Test}} & \textbf{Training} & \textbf{Model Size (Byte)} \\
		& \textbf{Accuracy} & \textbf{Accuracy} & \textbf{Time (s)} & \\
		\hline
		\textbf{QDA} & 94\% & 95.4\% & \cellcolor{green!30}12.447 & 5MB \\
		\textbf{Fine KNN} & \cellcolor{green!30}99.6\% & \cellcolor{green!30}100\% & 78.9 & 30MB \\
		\textbf{Cos KNN} & 90\% & 93\% & 86 & 30MB \\
		\textbf{Wide NN} & 86.5\% & 90.3\% & 50 & \cellcolor{green!30}296kB \\
		\hline
	\end{tabular}
	\caption{Performance Comparison of the four considered models in terms of validation and testing accuracy, training time and model size.}
	\label{tab:model_performance}
\end{table}

\input{table-res.tex}

\section{Conclusions}
In this work, we presented a pipeline for the efficient and effective classification of person-dependent SMRs using a large and public dataset. In particular, motor imagined movements (left and right) and rest phases have been successfully classified. The findings demonstrate the pipeline's promising capability to classify with accuracies regularly exceeding 99.5\%, ensuring reliable classification.
Despite the encouraging results, the approach will be validated using more datasets from the literature and a custom datasets in future works, to better corroborate the validity of the findings and draw final conclusions. Moreover, a cross-subject classification rather than a within-subject classification will be performed. If feasible, the cross-subject classification would facilitate the avoidance of long training phases for people, hence augmenting the likelihood of these systems being used in the real-world applications. Preliminary cross-subject testing has been conducted, and at the present moment, the current pipeline requires further modifications to enhance generalizability. These enhancements will be the focus of our future work. 


\bibliographystyle{IEEEtran}
\bibliography{./bibCustom}
\end{document}

%% file: symbols_def.tex
\usepackage{subcaption}
\usepackage{stix}
\usepackage{tikz, adjustbox}
\usepackage{pgfplots}
\usetikzlibrary{positioning, shapes.multipart}
\usepackage{ifthen}

\usetikzlibrary{shapes.geometric, arrows, positioning}
\tikzstyle{process} = [rectangle, minimum width=3cm, minimum height=1cm, text centered, draw=black, font=\bfseries]
\tikzstyle{subprocess} = [rectangle, minimum width=3cm, minimum height=1cm, text centered, draw=black, dashed]
\tikzstyle{arrow} = [thick,->,>=stealth]
\usetikzlibrary{3d,decorations.text,shapes.arrows,positioning,fit,backgrounds}
\tikzset{pics/fake box/.style args={
		#1 with dimensions #2 and #3 and #4 and label #5 and shift #6}{
		code={
			\draw[gray,ultra thin,fill=#1]  (0,0,0) coordinate(-front-bottom-left) to
			++ (0,#3,0) coordinate(-front-top-right) --++
			(#2,0,0) coordinate(-front-top-right) --++ (0,-#3,0) 
			coordinate(-front-bottom-right) -- cycle;
			\draw[gray,ultra thin,fill=#1] (0,#3,0)  --++ 
			(0,0,#4) coordinate(-back-top-left) --++ (#2,0,0) 
			coordinate(-back-top-right) --++ (0,0,-#4)  -- cycle;
			\draw[gray,ultra thin,fill=#1!80!black] (#2,0,0) --++ (0,0,#4) coordinate(-back-bottom-right)
			--++ (0,#3,0) --++ (0,0,-#4) -- cycle;
			\path[gray,decorate,decoration={text effects along path,text={}}] (#2/2,{2+(#3-2)/2},0) -- (#2/2,0,0);
			\node[anchor=south, align=center, xshift=#6cm ,yshift=-0.3cm] {\tiny #5};
		}
}}
\tikzset{circle dotted/.style={dash pattern=on .05mm off 2mm,
		line cap=round}}

\usepackage{accents}

\usepackage{color}

\newcommand{\vect}[1]{\boldsymbol{#1}}		
\newcommand{\nR}[1]{\mathbb{R}^{#1}}		
\newcommand{\upperRomannumeral}[1]{\uppercase\expandafter{\romannumeral#1}}	

\newcommand{\fig}{Fig.~}	
\newcommand{\eqn}{Eq.~}	
\newcommand{\tab}{Tab.~}	
\newcommand{\sect}{Sec.~}	



%% file: img/flow-chart.tex
\definecolor{myLightBlue}{RGB}{218,232,252}
\definecolor{myBlue}{RGB}{114,147,194}

\definecolor{myLightYellow}{RGB}{255,242,204}
\definecolor{myYellow}{RGB}{214,182,86}

\definecolor{myLightOrange}{RGB}{255,230,204}
\definecolor{myOrange}{RGB}{216,158,9}

\definecolor{myLightGreen}{RGB}{226,239,217}
\definecolor{myGreen}{RGB}{168,208,141}

\definecolor{myLightRed}{RGB}{255,220,204}
\definecolor{myRed}{RGB}{216,80,9}

\begin{figure*}[t]
	\centering
	\begin{adjustbox}{width=2\columnwidth}
	\begin{tikzpicture}[node distance=2cm]
	
	\node (data) [process, draw=white!60!myBlue, fill=white!60!myLightBlue] {Data  Import};
	\node (artifact) [process, right of=data, xshift=2cm, align=center, draw=white!60!myGreen, fill=white!60!myLightGreen] {Artifacts and\\ Channels Selection};
	\node (filtering) [process, right of=artifact, xshift=2cm, draw=white!60!myOrange, fill=white!60!myLightOrange] {Filtering};
	
	\node (featextr) [process, right of=filtering, xshift=2cm, text width=3.2cm, align=center, draw=white!60!myYellow, fill=white!60!myLightYellow] {Feature Extraction};
	
	\node (subdata1) [subprocess, below of=data, text width=3.2cm, align=center] {
		\begin{itemize}
		\item 52 Subjects~\cite{2017-HohAhnAhnKwoMooJun}
		\item 100 MI Trials (Left Right Hand, Rest)
		\end{itemize}
	};
	
	\node (subartifact1) [subprocess, below = 0.7cm of artifact, text width=3.2cm, align=left] {\begin{itemize}
		\item 10 Chs Selected
		\item Statistical Outlier Detection
		\end{itemize}};
	\node (subfiltering) [subprocess, below=0.7cm of filtering, text width=3.2cm, align=center] {		\begin{itemize}
		\item Band Pass: [7-30] Hz
		\item Common Average Reference (CAR)
		\end{itemize}};

		\node (subfeatextr) [subprocess, below=0.7cm of featextr, text width=3.2cm, align=center] {		\begin{enumerate}
		\item[1.] Energy
		\item[2.] Instantaneous Entropy
		\end{enumerate}};
		
	\node (scalp) [process, right=1cm of featextr, yshift=0cm, draw=white!60!myRed, fill=white!60!myLightRed] {Classification};
	\node (subscalp)[subprocess, below = 0.7cm of scalp, text width=3.8cm, align=center]{\begin{itemize}
		\item QDA
		\item Fine-KNN
		\item Cos-KNN
		\item Weighted-NN
		\end{itemize}};

	\draw [arrow] (data) -- (artifact);
	\draw [arrow] (artifact) -- (filtering);
	
	\draw [arrow] (data) -- (subdata1);
	
	\draw [arrow] (artifact) -- (subartifact1);

	\draw [arrow] (filtering) -- (subfiltering);
	
	\draw [arrow] (filtering.east) -- (featextr);
	
	\draw [arrow] (featextr) -- (subfeatextr);
	
	\draw [arrow] (featextr.east) -- (scalp);

	\draw [arrow] (scalp) -- (subscalp);

	\end{tikzpicture}
\end{adjustbox}
\caption{Flow Chart of the Methodology Applied. Data are collected from 52 subjects for a total of 100 trials per subject. Artifacts have been removed a statistical outlier detection. Ten Channels out of 64 have been considered. A bandpass filter and a CAR have been applied to the data. A feature extraction process, i.e. the computation of Energy and Instantaneous Entropy and a final classification stage conclude the procedure.}
\label{fig:framework}
\end{figure*}

%% file: table-res.tex
\begin{table}[t]
\centering
\resizebox{\columnwidth}{!}{%
\begin{tabular}{llllllllll|}
 \hline
\textbf{Approach} & \textbf{Accuracy} & \textbf{Recall} & \textbf{Specificity} & \textbf{$F_1$} \\  \hline
\cite{2017-HohAhnAhnKwoMooJun} \textbf{Cho et al.} & 0.6746 &  & &  \\
\cite{2019-Kum} \textbf{Kumar et al.}  & 0.6724 &  & &\\
\cite{2021-Sad} \textbf{Sadiq et al.} & 0.8502 & & & \\
\cite{2022-Sad} \textbf{Sadiq et al.}  & 0.8769 & 0.8762 & 0.8775 & 0.8770\\

	\cellcolor{green!30}\textbf{Ours} & \cellcolor{green!30}0.995 & \cellcolor{green!30}0.9986 & \cellcolor{green!30}0.9983 & \cellcolor{green!30}0.9984\\
  \hline
\end{tabular}%
}
\caption{Table summarizing the findings from four studies that analyzed the same dataset. Our results, which outperform the others, are highlighted in green.}
\label{tab:avg-res}
\end{table}